\documentstyle[prl,aps]{revtex}
\tighten
\begin{document}

\draft
\title{Cumulative Parity Violation In Supernovae}


\author{C. J. Horowitz\footnote{email: charlie@iucf.indiana.edu} 
and Gang Li\footnote{email: ganli@indiana.edu}} 
\address{Nuclear Theory Center\\
2401 Milo B. Sampson Lane \\
Bloomington, Indiana 47405}

\date{\today} 
\maketitle
 
\begin{abstract}
Supernovae provide a unique opportunity for large scale parity violation 
because they are dominated by neutrinos.  We calculate the parity violating 
asymmetry $A$ of neutrino emission in a strong magnetic field.  
We assume the neutrinos elastically scatter many times from slightly 
polarized neutrons.   Because of the multiple
interactions, $A$ grows with the optical thickness of the 
proto-neutron star and may be much larger than previous estimates.  
As a result, the neutron star could recoil at a significant velocity.  
\end{abstract}

Weakly interacting neutrinos dominate core collapse supernovae.  This 
provides a unique opportunity for large scale or macroscopic parity 
violation [1,2].  Therefore, it is of fundamental importance to study 
parity violation in a supernova.

We focus on an asymmetry induced in the explosion because of parity violation 
in a strong magnetic field.   This asymmetry could lead to a recoil of the 
newly formed neutron star.  Indeed, neutron stars have large velocities 
$\approx 500$ km/s [3].  An asymmetry of order one percent could produce 
these velocities [4]  (see below).   However, this asymmetry need not arise 
from parity violation (see for example [5]).

Strong magnetic fields are present in pulsars.  Indeed, external dipole 
fields of $10^{12}$ to $10^{13}$ G are inferred in many cases [6].   
In this paper, we estimate the magnitude of parity violating effects 
from {\it known} neutrino interactions in {\it known} fields of $10^{12}$ 
to $10^{13}$ G.  Others have speculated on parity violating effects in 
much stronger fields [1,2,7-11] and with new neutrino interactions [12].  
We find that repeated interactions as neutrinos diffuse through an 
optically thick medium may greatly enhance the asymmetry.

There is some controversy on observational correlations between recoil 
velocities and magnetic field strength.  For example, Birkel and Toldra [13] 
argue that for rapidly spinning pulsars there is no correlation  between the 
recoil velocity and the projection of B on the spin axis.   We think their 
analyses may be over simplified because they do not consider possible 
effects of rapid rotation on the dynamics of the collapse and on the 
asymmetry.  Unfortunately observational tests involve incomplete 
information.  The strength of the external dipole field is inferred.  
However, little is known about non-dipole fields.  We find that the most 
important variable may be the volume of the core occupied by the strong 
field (see below) rather than simply its strength.  
Thus we keep an open mind with respect to present observations.

Independent of observation, it is important to estimate parity violating 
effects.  We see three possibilities:   parity violating effects could be 
small and thus irrelevant,  they could be large and observed or they could 
be potentially large and not observed.  If they are not observed, it may 
still be possible to set useful limits on the magnetic field configuration 
or on new weak interactions.  In any case, we need accurate theoretical 
estimates.

Previous estimates of parity violation may be incomplete because they ignore 
possible enhancements from repeated interactions.   Neutrino transport 
involves diffusion with neutrinos undergoing many parity violating 
interactions before they escape.  We estimate these cumulative effects below.

Much previous work focused on electrons [1,11].  It is natural to think that 
neutrino electron scattering will dominate the asymmetry because the 
electron's magnetic moment is 1000 times that of a nucleon.   However, the 
electron polarization is reduced because they are relativistic and 
degenerate.  Furthermore, because of the small $\nu-e$ cross section this 
polarization may lead to an asymmetry that is {\it smaller} than that from 
nucleon reactions.  It is important to examine other processes to identify 
the largest contribution to the asymmetry.

In this paper, we consider neutrino elastic scattering from slightly polarized 
neutrons.  This may be important (even though the neutron 
polarization is small).  We do not claim that it is the largest contribution.  
Instead we focus on elastic neutron scattering for simplicity.  
The differential cross section is (see for example [14]),
$$d\sigma/d\Omega = {G^2E_\nu^2\over 4\pi^2}\bigl\{c_v^2+3c_a^2
+(c_v^2-c_a^2){\rm cos}\theta
+ 2P_nc_a[(c_v-c_a){\rm cos}\theta_{in}+
(c_v+c_a){\rm cos}\theta_{out}]\bigr\}.\eqno(1)
$$
Here $G$ is the Fermi constant, $E_\nu$\ the neutrino energy (assumed much 
smaller than the nucleon mass $M$) and $c_v=-1/2$, $c_a=-g_a/2$ with 
$g_a=1.26$.  The incident neutrino momentum  makes an angle $\theta_{in}$ 
with the polarization direction,  scatters through an 
angle $\theta$ and then the outgoing momentum is at an angle $\theta_{out}$ 
with the polarization.

The polarization of the neutrons $P_n$ depends on the magnetic field $B$ and 
temperature $T$, $P_n\approx eB/ M T$,
$$P_n\approx 2\times 10^{-5} 
[{B\over 10^{13} \ {\rm G}} ] [{{\rm 3\ MeV}\over T} ].\eqno(2)$$ 
In the limit $g_a=1$\ Eq. (1) becomes,
$$d\sigma/d\Omega = \sigma_0 ( 1 + P_n {\rm cos}\theta_{out}),\eqno(3)$$
with $\sigma_0=G^2E_\nu^2/4\pi^2$.  This simple form provides insight and 
is good to 10 percent for the asymmetry.  [However, we use the full result Eq. 
(1) in the Monte Carlo below.]   Equation (3) does not depend on 
$\theta_{in}$.  Thus the mean free path is independent of direction.  The 
asymmetry arises because the outgoing neutrino angular distribution is biased 
towards the polarization direction.

The total dipole asymmetry $A$ in the neutrino angular distribution 
$I(\theta)$ is, 
$$A=\int_{-1}^1 d{\rm cos}\theta\ I(\theta) \ {\rm cos}\theta / 
\int_{-1}^1 d{\rm cos}\theta\  I(\theta).\eqno(4)$$
If the angular distribution from the supernova is proportional to Eq. (3) then 
$A=P_n/3$.  This asymmetry is  related to the recoil velocity of the star.

The gravitational binding energy of a neutron star is of order 100 MeV per 
nucleon.  This is radiated away in neutrinos of momentum 100 MeV/n leaving 
a proto-neutron star of mass about 839 MeV/n.  Therefore the recoil velocity 
$v$ of the star is,
$$v/c\approx {100\over 839} A \ \approx 0.1 A.\eqno(5)$$
Thus if $A$ is only of order $P_n$  the velocity will be small 
(around  $10^{-6}$ of the speed of light $c$ for $B$ near $10^{13}$ G).

Neutrinos must diffuse through many mean free paths in order to escape the 
star so they interact repeatedly with the polarized neutrons.  The crucial 
question is do these repeated interactions enhance the asymmetry?   If the 
temperature distribution at the neutrino sphere is independent of direction 
than what happens inside may not be important.  The asymmetry in the 
neutrino flux will simply arise from the last scattering and be of order $P_n$.

However, neutrinos dominate the energy transport.  [Note, the effects of 
convection on $A$ remain to be investigated.]  Therefore we expect the 
temperature distribution to be asymmetric because of the asymmetric 
neutrino flux.  This could lead to an asymmetry much larger than $P_n$.  To 
investigate multiple interactions, we calculate the asymmetry of a  
``reference configuration''  with a very simple Monte Carlo.

This reference configuration is not meant to be a realistic supernova 
simulation.  Instead it is the simplest system with slight nucleon 
polarization and repeated neutrino interactions.  This may allow a simple 
exploration of the physics.  We consider a uniform sphere of slightly 
polarized neutrons.  The neutrinos start either at the center or uniformly 
throughout the volume and then interact only via elastic neutron scattering, 
Eq. (1).  We discuss these assumptions below.

The model has two parameters: the polarization of the neutrons $P_n$ and the 
optical depth of the sphere $r/\lambda$.  This is the radius $r$ measured in 
units of the neutrino mean free path $\lambda$.  For $B=5\times 10^{12}$ G 
and $T=3$ MeV the polarization is
$$P_n\approx 1\times 10^{-5}.\eqno(6)$$
Near the center of the star the temperature is larger and the neutrons will 
become slightly degenerate.  This will decrease $P_n$ somewhat.  Also at high 
densities, strong interactions may modify $P_n$.  It is even possible that 
there is a ferromagnetic phase of dense neutron rich matter [15].   (While 
this is unlikely at the high temperatures of a supernova, there could still 
be an enhancement in $P_n$.)  Perhaps most importantly, we are assuming that 
the strong $B$ field penetrates the central region.  If $B$ is excluded, 
the average polarization will be lower.  Alternatively,  there could be 
regions of very high internal fields.  For simplicity, we adopt Eq. (6) 
for $P_n$ and assume that it is uniform throughout the sphere.  Of course, 
the final asymmetry is proportional to $P_n$ so it is easy to consider other 
values.

We chose the optical depth so that the time scale for neutrinos to diffuse 
out of the sphere is approximately correct.  Neutrinos were detected from 
SN1987A over about 10 seconds [16].  The escape time $t$ is of order the 
light travel time $r/c$ ($\approx 0.1$ msec) multiplied by the optical depth 
$r/\lambda$.  For $t$ to be of order one second requires,
$$r/\lambda \approx 1\times 10^4.\eqno(7)$$
This value is consistent with theoretical simulations [17].

We have calculated the overall asymmetry, Eq. (4) using a very simple Monte 
Carlo code, see Figure 1.  To save computer time we use a larger $P_n=0.01$\ 
than Eq. (6) and smaller $r/\lambda$ than Eq. (7).   The 
results can be scaled to the desired values.  We find that 
{\it the asymmetry grows with $r/\lambda$},
$$A\approx \alpha P_n ({r\over \lambda}) + O(P_n{r\over 
\lambda})^2.\eqno(8)$$
This is our most important result.  Note, the second term in Eq. (8) is
needed since $A$ saturates for very large $r/\lambda$.  
The coefficient $\alpha$ arises primarily from the 
angle averaging in the Monte Carlo.  For an r=0 source $\alpha\approx 
0.14$.   For uniformly distributed sources $\alpha$ is somewhat 
smaller $\alpha\approx  .057$.  This reflects the shorter path length for 
neutrinos starting close to the surface.  However, $A$ still grows strongly 
with $r/\lambda$.

The linear dependence of Eq. (8) on $r/\lambda$ can be understood with a 
simple one dimensional biased random walk.  At each time step the probability 
to hop left is $0.5 + P_n/2$ (and right is $0.5-P_n/2$).  Because of this 
bias, the mean value of the neutrino's  position $<x>$ is not zero but 
drifts left with a velocity of order $P_n$.  Therefore $<x>$ is proportional 
to $t$.  In contrast, the width of the neutrino distribution grows because 
of diffusion but only with $t^{1/2}$.  The macroscopic asymmetry $A$ depends 
on the ratio of the mean value to the width.  Therefore $A$ grows with 
$t/t^{1/2}=t^{1/2}$.  Finally, the time to escape $t$ is proportional to 
$(r/\lambda)^2$ so $A$ grows linearly with $r/\lambda$.

For the parameters of Eqs. (6-7) the reference configuration asymmetry is
$$A\approx 0.014,\eqno(9)$$
(assuming an r=0 source).  This $A$  would imply a recoil of the proto-neutron 
star of $\approx 400$ km/s.    This value is interesting and much larger 
than previous estimates.  However, it is only a ballpark estimate and we must 
discuss a number of our assumptions.

First, we only considered neutrino-neutron elastic scattering.  For mu and tau 
neutrinos this is a reasonable first approximation to the opacity.  However, 
pair production and annihilation can change the number of neutrinos.  Multiple 
interactions will tend to bring neutrinos into thermodynamic equilibrium with 
the other matter.  The number of neutrinos then depends on the temperature.  
Thus, the asymmetric neutrino flux will lead to an asymmetric temperature.   
One side of the star will be warmer than the other side.  We expect the 
asymmetry in the temperature to be small near r=0 and grow as one moves out 
nearer the neutrino sphere.  Microscopic simulations to study the temperature 
distribution would be very useful.

For electron neutrinos one should also consider neutrino capture followed by 
electron capture.  These reactions also have asymmetries.  For electron 
capture the asymmetry depends on the electron polarization $P_e$,  which can 
be somewhat higher than for neutrons.  However, $P_e$ is multiplied by the 
small coefficient  $(c_v^2-c_a^2)/(c_v^2+3c_a^2)\approx 0.1$ [7,9] so its
contribution may be similar to Eq. (1).

It is important to discuss antineutrinos.  The antineutrino-neutron 
elastic cross section is related to Eq. (1) by switching cos$\theta_{in}$ 
with cos$\theta_{out}$,
$$d\sigma/d\Omega = {G^2E_\nu^2\over 4\pi^2}\bigl\{c_v^2+3c_a^2
+(c_v^2-c_a^2){\rm cos}\theta
+ 2P_nc_a[(c_v+c_a){\rm cos}\theta_{in}+
(c_v-c_a){\rm cos}\theta_{out}]\bigr\}.\eqno(10)
$$      
This will lead to an opposite sign for the asymmetry.  Thus there will be 
some cancellation between neutrinos and antineutrinos.  However, this 
cancellation is not perfect because the opacity is different for electron 
$\nu$ and $\bar\nu$.  Note, the physical mechanism is different for
$\nu$ and $\bar\nu$.  For $\nu$, the mean free path is almost 
independent of direction but the outgoing angular distribution is 
biased. For $\bar\nu$, Eq. (3) becomes $d\sigma/d\Omega=\sigma_0 
(1+P_n{\rm cos}\theta_{in})$ so that the mean free path depends on direction
but the outgoing angular distribution is almost unbiased.

Neutrino and antineutrino asymmetries will {\it add} in calculating the 
net lepton number flux.  This could lead to significant asymmetries in the 
trapped lepton fraction and in the chemical composition.  If the 
individual fluxes are say 10 times the net lepton number flux then this 
may be very sensitive to small asymmetries.
One percent changes in the $\nu$ and $\bar\nu$ fluxes 
will lead to a twenty percent asymmetry in the lepton 
number flux.  This may have an important impact on the explosion.  For 
example, on one side of the star the larger neutrino flux may enhance the 
proton fraction while on the other side the antineutrino flux may increase the 
number of neutrons.  One should investigate asymmetries in the chemical 
composition.

In this paper we examined the asymmetry induced in a supernova from 
repeated parity violating neutrino-neutron elastic scatterings in a strong 
magnetic field.  As a model, we calculated the asymmetry of neutrinos 
emerging from an optically thick sphere of slightly polarized neutrons.  We 
find that the asymmetry {\it grows} with the optical thickness through 
which the neutrinos diffuse.  This is because the neutrinos interact very 
many times and leads to parity violating effects that are much larger than 
in previous estimates.

Future work should examine asymmetries induced in the temperature and 
chemical composition of the proto-neutron star.  One should also examine 
asymmetries in other reactions and the magnetic properties of very dense 
neutron rich matter.   We note that parity violation may be sensitive to 
conditions deep inside the proto-neutron star in addition to near the neutrino 
sphere.  In particular, the volume of the core occupied by a strong magnetic 
field may influence the asymmetry.

This work was supported in part by DOE grant: DE-FG02-87ER40365.

\eject

\vbox to 4.in{\vss\hbox to 8in{\hss
{\includegraphics{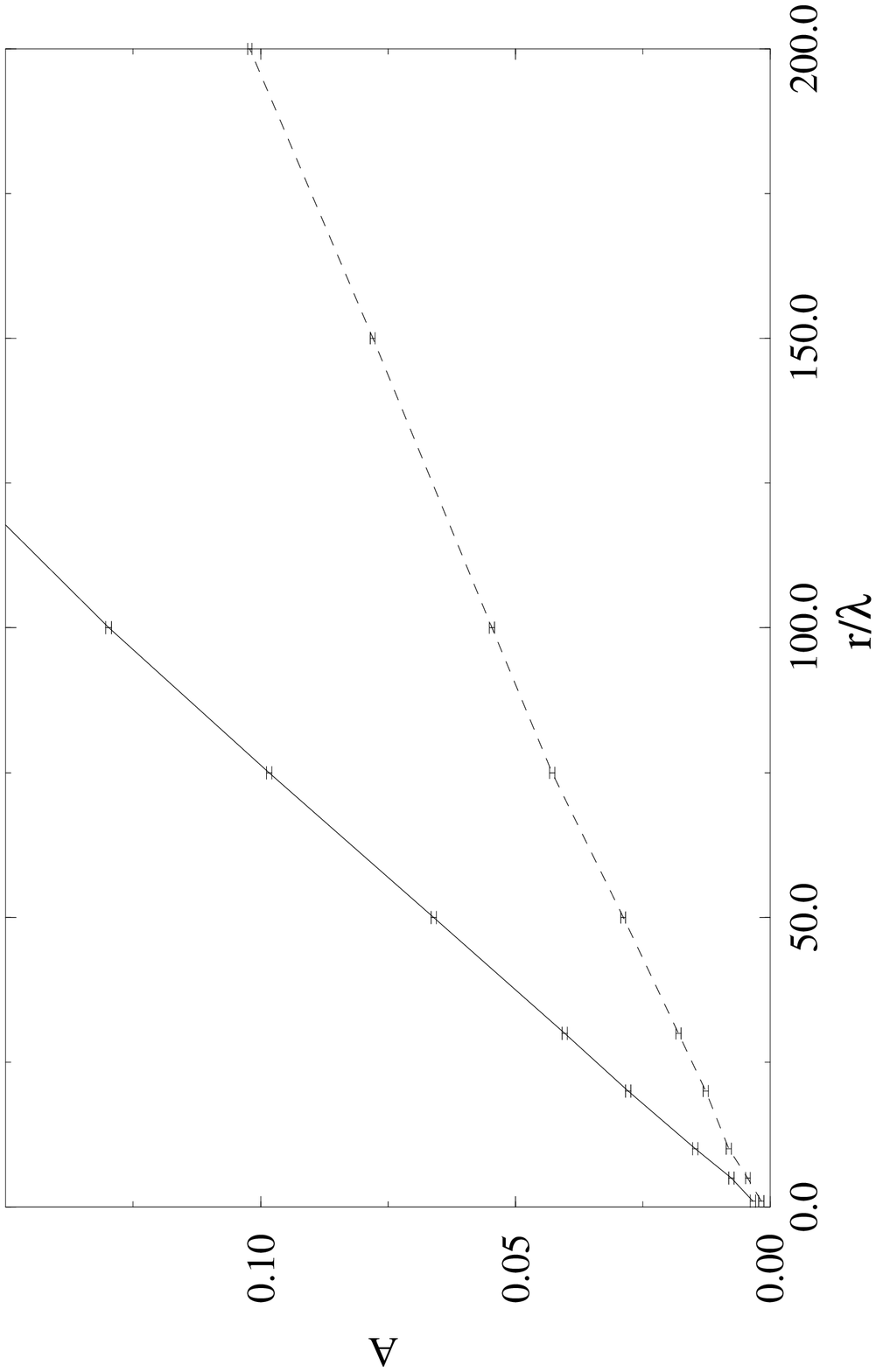}}\hss}}
\nobreak
{\noindent\narrower{{\bf FIG.~1}. 
Asymmetry $A$, Eq. (4), of neutrinos emitted from a neutron sphere of optical
depth $r/\lambda$.  The polarization of the neutrons is $P_n=0.01$. The solid
curve assumes the neutrinos start at $r=0$ while the dotted curve is for 
neutrinos starting uniformly throughout the volume.}}


\begin{references}
\bibitem{} A. Vilenkin, 1979, unpublished; 
Ap J {\bf 451} (1995) 700. 

\bibitem{} N. N. Chugai, Pisma Astron. Zh. {\bf 10} (1984) 210. 
[Sov. Astron. Lett. {\bf 10} (1984) 87.]

\bibitem{} A. G. Lyne and D. R. Lorimer, Nature {\bf 369} (1994) 127.

\bibitem{}  S. E. Woosley in ``The Origin and Evolution of Neutron Stars,'' 
eds. D. J. Helfand and J. H. Huang (D. Reidel:Dordrecht)  p. 255 (1987).

\bibitem{} A. Burrows and J. Hayes, Phys. Rev. Lett. {\bf 76} (1996) 352.

\bibitem{} V. S. Beskin, A. V. Gurevich, and Ya. N. Istomin, 
Physics of the Pulsar Magnetosphere, Cambridge University Press,
Cambridge, 1993.

\bibitem{} O. F. Dorafeev, V. N. Rodionov, and I. M. Ternov, Sov. Astron. 
Lett. {\bf 11} (1985) 123.

\bibitem{} R. C. Duncan and C. Thompson, Ap J
{\bf 392} (1992) L9.


\bibitem{} C. J. Horowitz and J. Piekarewicz, Los Alamos preprint 
archive, hep-ph/9701214.


\bibitem{} A. V. Kuznetsov and N. V. Mikheev, Phys. Lett. {\bf B394} (1997) 
123.
 
\bibitem{} V. G. Bezchastnov and P. Haensel, Phys. Rev. {\bf D54} (1996) 3706.

\bibitem{} A. Kusenko and G. Segr\'e, Phys. Rev. Lett. {\bf 77} (1996) 4872. 
E. Kh. Akhiredov, A. Lanza, and D. W. Sciama, hep-ph/9702436.

\bibitem{} Michael Birkel and Ramon Toldra, Los Alamos preprint archive, 
astro-ph/9704138.

\bibitem{} C. J. Horowitz and K. Wehrberger, Nucl. Phys. {\bf A531} (1991)665.


\bibitem{} M. Kutschera and W. Wojcik, Phys. Lett. {\bf B223}  (1989) 11.

\bibitem{} K. Hirata et al., Phys. Rev. Lett. {\bf 58} (1987) 1490;
R. M. Bionta et al., Phys. Rev. Lett. {\bf 58} (1987) 1494.

\bibitem{} Adam S. Burrows in ``Supernovae'' ed. A. G. Petschek, 
Springer-Verlag (N.Y. 1990) p143.
\end{references}
\end{document}